\documentclass[10pt,letterpaper,twocolumn]{revtex4-1} %% two column, final layout

\usepackage{graphicx}
\usepackage{amsmath}

\begin{document}

%\twocolumn[ %% activate for two-column option

\title{Whispering Gallery Mode Resonator Stabilized Narrow Linewidth Fiber Loop Laser}

%% For REVTeX it is possible to automate superscript and e-mail callouts with the superscriptaddress option; see REVTeX4 documentation.

\author{B. Sprenger, H. G. L. Schwefel, and L. J. Wang$^{*}$}

\address{
Max-Planck-Institute for the Science of Light\\
G\"unther-Scharowsky-Stra{\ss}e 1, Bau 24, 91058 Erlangen, Germany\\
$^*$Corresponding author: +49(9131) 6877-200, Lijun.Wang@mpl.mpg.de
}

\begin{abstract}We demonstrate a narrow line, fiber loop laser using Erbium-doped fiber as the gain material, stabilized by using a microsphere as a transmissive frequency selective element. Stable lasing with a linewidth of 170~kHz is observed, limited by the experimental spectral resolution. A linear increase in output power and a red-shift of the lasing mode were also observed with increasing pump power. Its potential application is also discussed.\end{abstract}

%\ocis{060.2410, 130.7408, 140.3425, 140.3510.}

%] %% activate for two-column option
\maketitle
%\doublespacing

\noindent Whispering gallery mode (WGM) resonators have been of great interest due to their small sizes and their very high Q factors \cite{vahala2003}. The most common types are microspheres \cite{chang1986,braginsky1989}, on-chip microdisks \cite{little1997}, crystalline microdisks \cite{grudinin2006}, and bottle-neck resonators \cite{pollinger2009}. Being small in diameter, $50-100~\mu$m, WGM resonators can have a very large free spectral range of several nanometers, offering a unique combination of single-mode transmission while maintaining a high quality factor.

Due to their high Q factors and compact size WGM resonators have the potential to be used as etalons for laser frequency stabilization, which are useful for telecommunications and frequency metrology \cite{matsko2007}. Fiber lasers have many advantages over discrete solid state or other types of lasers, including their long lifetimes and low maintenance since no further alignment is necessary. They also give almost ideal Gaussian modes with M$^2$ values of close to 1 with compact size, and can operate with air-cooling as opposed to water-cooling requirements in high-power lasers.

In the past WGM lasers have been created out of droplets with dye or quantum dots \cite{schafer2008}, sol-gel infused silica spheres \cite{yang2005}, and rare-earth doped glass spheres \cite{miura1996}. Some experiments have been conducted using passive whispering gallery modes for laser stabilization, although these require additional frequency selective elements \cite{kieu2006,kieu2007}. Travelling wave ring resonators are known to be more stable. An advantage of frequency selection by a passive resonator is the lack of absorption. A pure silica resonator, as described in this Letter, has very little absorption and allows for Q factors up to $10^{9}$ \cite{braginsky1989}, and thus has good potential for extremely high finesse laser stabilization. Experiments using crystalline disks are also planned, promising even higher passive Q factors of up to $10^{11}$ due to the low absorption \cite{grudinin2006}.

In this Letter, we report the observation of the very narrow linewidth lasing from a fiber loop laser stabilized with a WGM resonator. Each roundtrip in the loop laser only allows narrow modes to be transmitted and thus receive gain. The transmission is coupled into the microsphere using a tapered fiber, and out using an angle-polished fiber. The resulting lasing spectrum is studied with an optical spectrum analyzer and a heterodyne beat measurement. The resolution limit of both methods is reached. A red-shift in emission is seen with increasing pump power and is discussed.

\begin{figure}[tb]
\centering
\includegraphics[width=8.4cm]{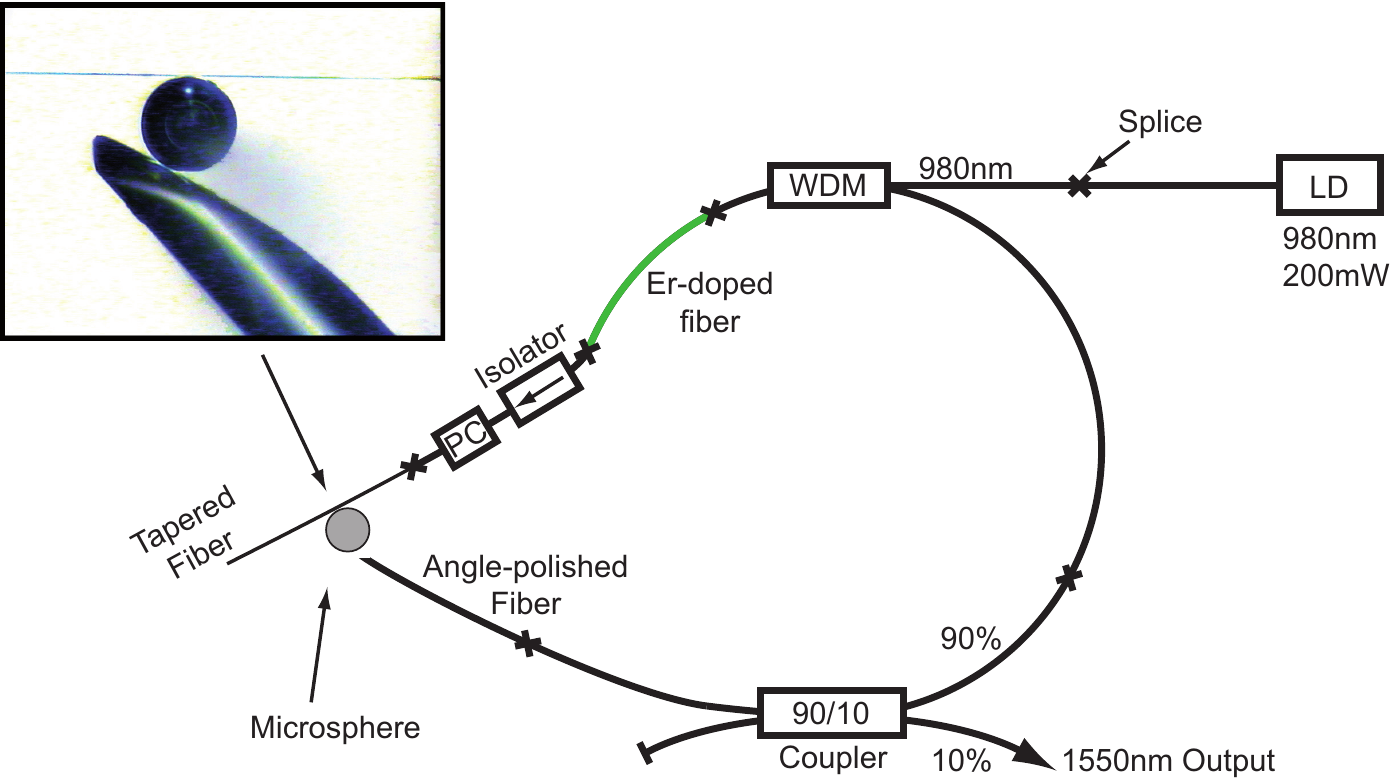}
\caption{(Color online) Stabilized fiber loop laser setup. A DFB laser (LD) pumps the Erbium loop laser through a wavelength division multiplexer (WDM). An isolator prevents backwards lasing, and polarization is controlled (PC) before coupling into a microsphere using a tapered fiber, and out using an angle-polished fiber. 10\% of the 1550~nm emission is coupled out. Inset, microscope image of the taper, microsphere, angle-polished fiber configuration.}\label{fig:SetupFigure}
\end{figure}

The microspheres in this experiment were created by melting the tips of tapered telecom fibers using a CO$_2$ laser. Sizes studied were between 50 and 100~$\mu$m in diameter, allowing for a free spectral range between 6 and 12~nm. A single-mode 1550~nm fiber tapered to a $1-2~\mu$m diameter \cite{knight1997} was used to couple the light into the microsphere, as shown in Fig.~\ref{fig:SetupFigure}. The large evanescent field of such a tapered fiber can be efficiently overlapped with the evanescent field of the high Q resonances of the microsphere. Critical coupling was achieved by carefully adjusting the position of the microsphere with respect to the tapered fiber using a three-dimensional piezoelectric stage. The light was coupled out using a fiber polished at a sharp angle of about $74^{\circ}$ as calculated for a 100~$\mu$m sphere \cite{ilchenko1999}. Total internal reflection due to the large angle causes an evanescent field outside such an angle-polished fiber. The fiber was also fixed on a piezoelectric stage to allow precise overlapping of the evanescent fields of the fiber and the microsphere. In this configuration, the narrow transmission bands of the microsphere are used as a frequency selective element inside the lasing cavity.

A fiber loop laser was set up using 70~cm of Erbium-doped fiber and an isolator to prevent lasing in the backward direction. A 200~mW DFB pump laser at 980~nm was coupled in using a wavelength division multiplexer. 10\% of the circulating power was coupled out with a fiber coupler, and multi-mode emission was observed around 1560~nm. The 10\% out-coupling helped to adjust the in-coupling into the angle-polished fiber, although during lasing the output of the tapered fiber was identical and even more powerful, so in a finished configuration the 10\% coupler is unnecessary. The fiber loop laser had peaks with linewidths around 0.2~nm, corresponding to 25~GHz, as measured using an optical spectrum analyzer (Agilent 86142). The length of the entire loop measured 7~m, which resulted in a very small mode separation of around 30~MHz. Generally two to four modes would lase simultaneously in an uncontrollable manner, in a range of $\sim 5$~nm. Opening the loop and inserting a microsphere, tapered fiber, and angle-polished fiber, as shown in Fig.~\ref{fig:SetupFigure}, only allow very specific and narrow modes to lase, and thus stabilize the fiber laser.

\begin{figure}[tb]
\centering
\includegraphics[width=8.4cm]{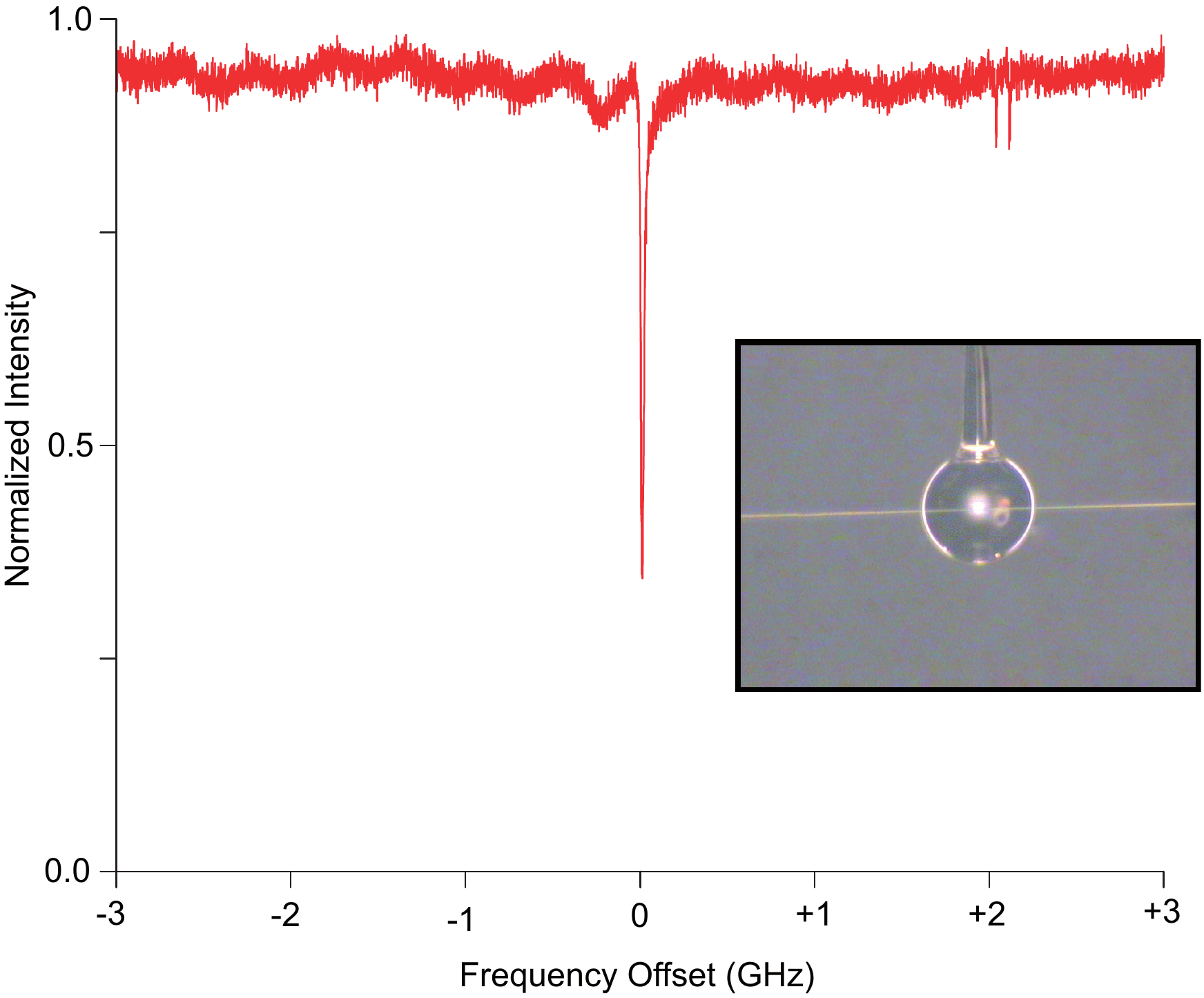}
\caption{Transmission of tapered fiber coupling to a microsphere. A grating stabilized diode laser is scanned across the resonance, scanning step 0.6~MHz. The Q factor is about 10$^7$. Inset, microscope image of microsphere coupling to taper.}\label{fig:ColdCavity}
\end{figure}

To measure the Q factor of the cavities we used a homebuilt grating stabilized diode laser at 1550~nm in the Littrow configuration \cite{hansch1972}. Light was coupled into the tapered fiber, and the intensity of the transmission was measured. The microsphere was brought towards critical coupling, and the frequency of the laser was tuned linearly. On resonance the photon lifetime in the cavity increased drastically, leading to a drop in intensity of 70\% or more. Figure~\ref{fig:ColdCavity} shows such a resonance with a linewidth of 17~MHz, corresponding to a Q factor of 10$^7$.

Precise positioning of the angle-polished fiber and the microsphere with respect to the tapered fiber allowed a laser cavity to be formed. Stable lasing of two to three modes was achieved. It was possible to get these modes to lase at similar intensities, and separations of up to 6~nm of two lasing modes was induced. Further adjustment of the in-going polarization as well as the coupling resulted in single-mode lasing. By including a thermoelectric cooling element and a PID controller the single-mode lasing peak position remained constant over 10 hours in a long-term measurement using a wavemeter (Advantest Q8236), and was below the resolution limit of 100~MHz. The optical to optical efficiency of the laser was about 0.1\%, as compared to 2\% in a closed loop configuration. This is due to the loss through the taper and coupling into and out of the microsphere, totaling about 6~dB loss per roundtrip.

\begin{figure}[tb]
\centering
\includegraphics[width=8.4cm]{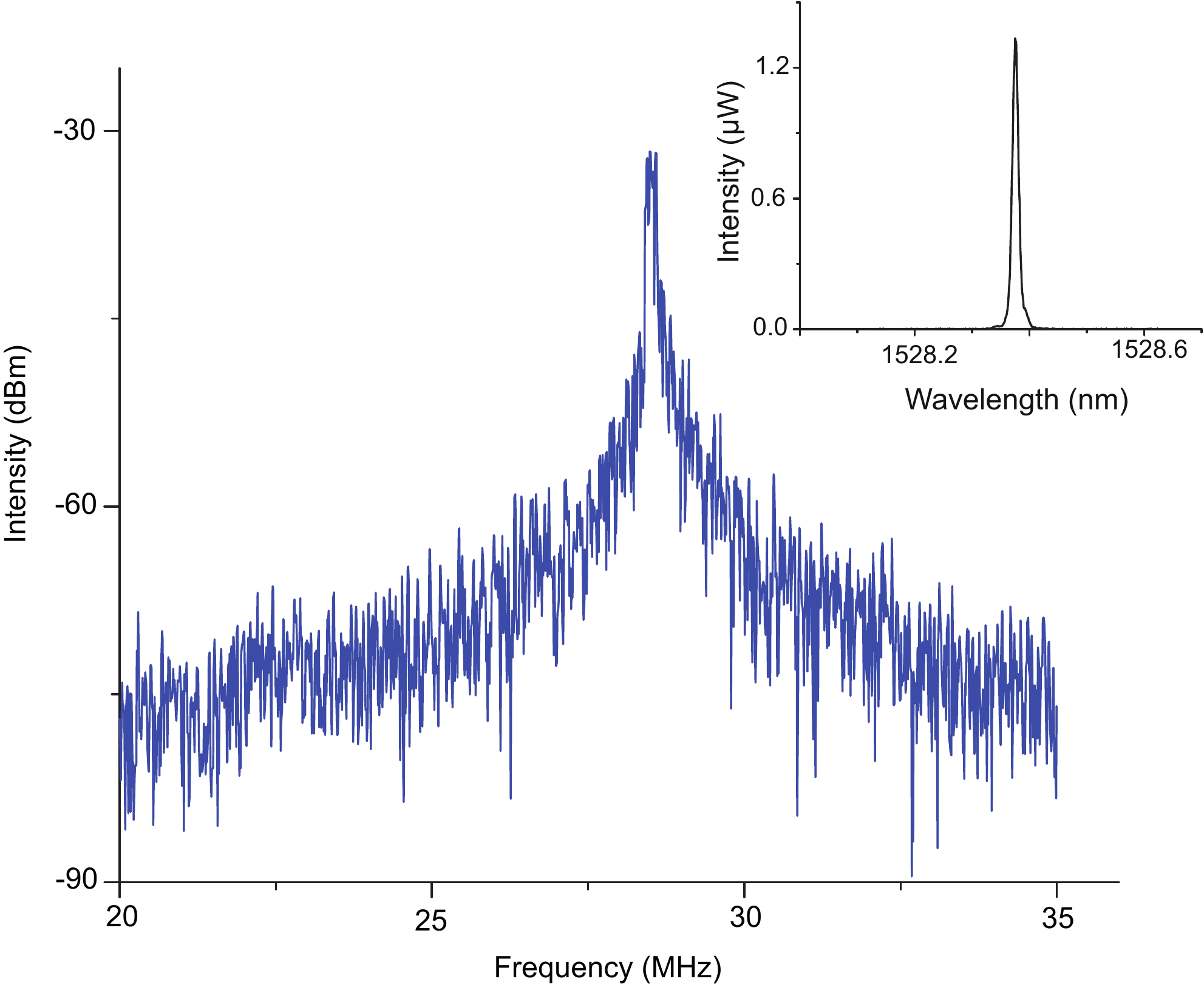}
\caption{Heterodyne beat signal between a grating stabilized diode laser and the stabilized fiber laser. The 3~dB linewidth of 170~kHz corresponds to a Q factor of $10^{9}$. Inset, stable lasing observed with an optical spectrum analyzer with a resolution of 0.01~nm.}\label{fig:LaserOutput}
\end{figure}

The linewidth measured with an optical spectrum analyzer (Ando AQ6317B) was resolution-limited at 0.01~nm, shown in the inset of Fig.~\ref{fig:LaserOutput}. For more precise determination of the linewidth a heterodyne beat measurement was performed, using the grating stabilized diode laser as a reference. By tuning the reference laser a few MHz away from the stabilized fiber laser, and overlapping the beams of the lasers on a fast photo-diode, an electronic beat signal could be measured showing the combined linewidths. We achieved a 170~kHz linewidth, corresponding to a Q factor of $10^{9}$ as shown in Fig.~\ref{fig:LaserOutput}. Similar results have been obtained using a microsphere as a reflective element in a different configuration \cite{kieu2006}. Since this measured linewidth is very close to the approximate linewidth of the reference laser, we are at the resolution limit of the measurement. Using cavities with Q factors of 10$^8$ or 10$^9$ we expect the true lasing linewidth to be even less. Further measurements using a frequency comb reference technique are planned.

\begin{figure}[h]
\centering
\includegraphics[width=8.4cm]{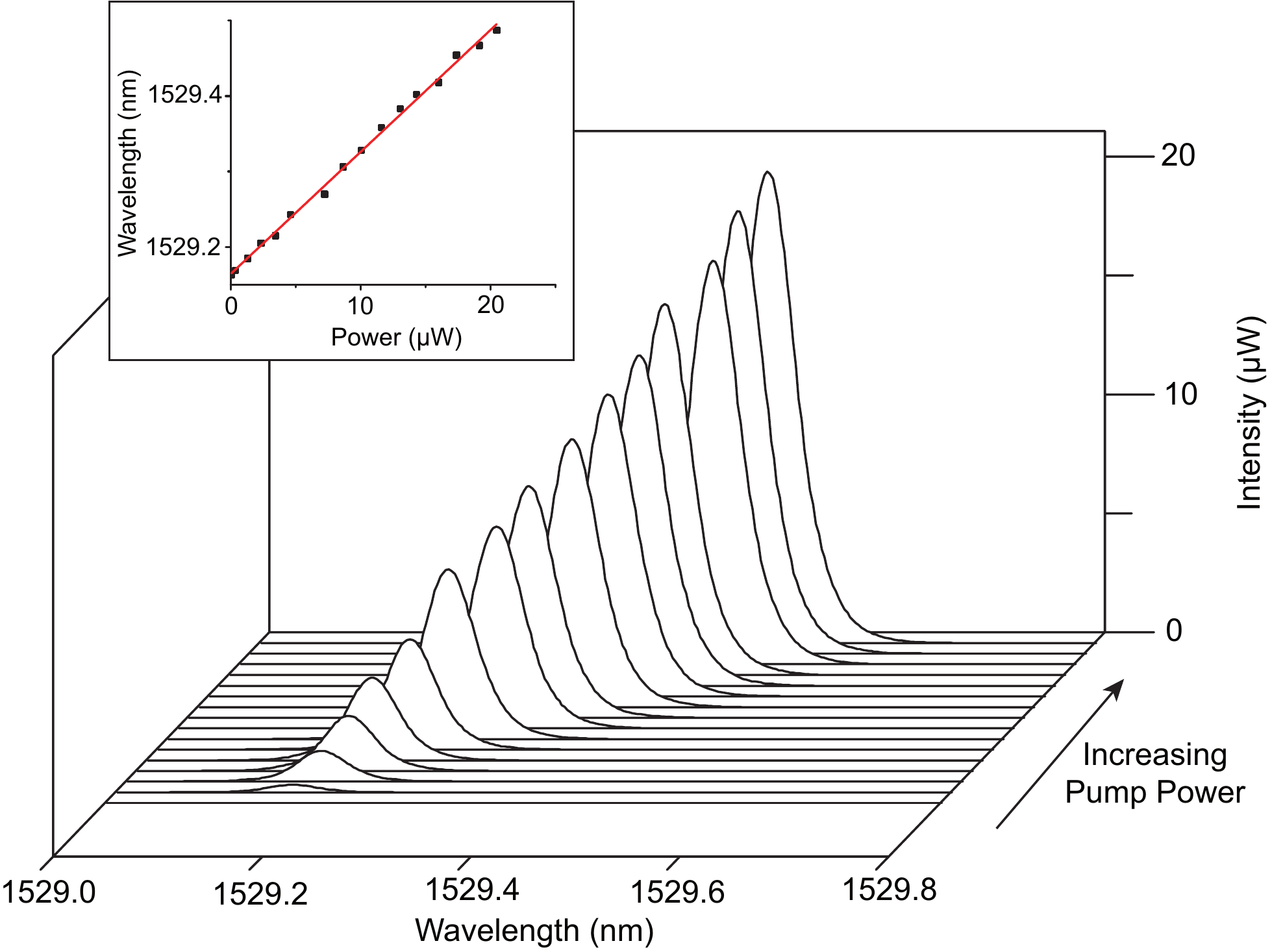}
\caption{Pump power dependence of lasing emission shows a linear increase in output intensity, as well as a red-shift. Inset, wavelength as a function of output power showing the linear red-shift.}\label{fig:PumpPower}
\end{figure}

Figure~\ref{fig:PumpPower} shows that the lasing output power increases linearly with pump power of the 980~nm laser. Furthermore, a red-shift can be observed with increasing pump power. This corresponds to the expected thermo-refractive index at high powers inside the cavity. The photon lifetime in the cavity is very long and is partly limited by absorption, which leads to heating along the path. Since $\tilde{n} = n_0 + n_2~I + (dn/dT)~\tilde{T}$, and in fused silica the thermo-refractive index plays a larger role than the $\chi^{(3)}$ nonlinearity \cite{boyd1992,matsko2007}, as well as the thermal expansion  of the microsphere's circumference, we are only concerned with the $(dn/dT)~\tilde{T}$ term. These three terms all lead to an increase in the refractive index of the material at increasing intensity. The slope of the red-shift as a function of output power is 16~pm/$\mu$W and can be seen in the inset of Fig.~\ref{fig:PumpPower}. With a pump power of 200~mW at 980~$nm$, and 10\% out-coupled power of 20~$\mu$W the refractive index changes by 0.00031. The $(dn/dT)$ term for fused silica is $1.2\times 10^{-5}K^{-1}$ \cite{boyd1992}, resulting in a maximum induced temperature change of 25.8~K. Since the lasing frequency is primarily determined by the microsphere, whose mass is very small, rapid tuning and further stabilization of frequency is achievable by varying the pump intensity.

In a ``bad-cavity'' laser \cite{kuppens1994,exter1995} the cavity loss rate $\Gamma_0$ is comparable to or larger than the gain bandwidth. In the case of this laser the gain bandwidth can be thought of as the transmission bandwidth through the microsphere, which is just over $10^7$~Hz. The cavity loss rate $\gamma$ is very high at several $10^7~s^{-1}$ as well, so we are entering the bad-cavity regime. It is known that the Schawlow-Townes laser linewidth limit does not apply in this case, so very low linewidths should be possible, and will be further investigated.

In conclusion, we demonstrate the stabilization of a fiber loop laser by forcing transmission through a high Q whispering gallery mode resonator, using a tapered fiber as well as an angle-polished fiber. The WGM microsphere is the only wavelength selecting element used in the setup, allowing different resonances to be used within the gain spectrum. Any type of gain material could be used inside the fiber with a corresponding pump laser. We achieved single-mode lasing that could be kept constant over hours with less than 100~MHz drift. The linewidth was determined to be 170~kHz or less, limited by the experimental resolution. This corresponds to a Q factor of at least $10^{9}$. A linear increase in output power, as well as a red-shift were observed with increasing pump power. Lasing Q factors were at least two orders of magnitude higher than the passive Q factors of the cavities, so crystalline whispering gallery mode disks with Q factors of 10$^{11}$ will allow even more precise frequency sources. Further measurements using a frequency comb should allow for a more precise determination of the linewidth and the Q factor. We believe that the method demonstrated in this Letter offers a genuine way for stabilizing fiber lasers at all wavelengths, and may offer important applications to various fields such as frequency metrology.

The authors would like to thank Jie Zhang, Zehuang Lu, Vladimir Elman, and Sascha Preu for stimulating discussions.
%\pagebreak

%\bibliographystyle{osajnl}
%\bibliography{Bibliography}

\end{document}